# MAGNET: Understanding and Improving the Accuracy of Genome Pre-Alignment Filtering

Alser, Mohammed; Mutlu, Onur; and Alkan, Can

**Abstract:** *In the era of high throughput DNA sequencing (HTS) technologies, calculating the edit distance (i.e., the minimum number of substitutions, insertions, and deletions between a pair of sequences) for billions of genomic sequences is the computational bottleneck in today's read mappers. The shifted Hamming distance (SHD) algorithm proposes a fast filtering strategy that can rapidly filter out invalid mappings that have more edits than allowed. However, SHD shows high inaccuracy in its filtering by admitting invalid mappings to be marked as correct ones. This wastes the execution time and imposes a large computational burden. In this work, we comprehensively investigate four sources that lead to the filtering inaccuracy. We propose MAGNET, a new filtering strategy that maintains high accuracy across different edit distance thresholds and data sets. It significantly improves the accuracy of pre-alignment filtering by one to two orders of magnitude. The MATLAB implementations of MAGNET and SHD are open source and available at: https://github.com/BilkentCompGen/MAGNET.*

## 1. INTRODUCTION

Until today, it remains challenging to sequence the entire DNA molecule as a whole. As a workaround, High throughput DNA sequencing (HTS) technologies are used to sequence random fragments (called *short reads*, which are 75-300 base-pairs long) of copies of the original molecule. The biggest challenge with these technologies is the use of these short reads to construct the *complete* genome sequence (~3.2 billion base-pairs for human genome), as these reads do not have any information about which part of genome they come from. During this process, called *read mapping*, each read is mapped to a reference genome based on the similarity between the read and "candidate" locations in that reference genome (like solving a jigsaw puzzle). The similarity measurement, called *alignment* or *verification*, is formulated as an approximate string matching problem and solved using quadratic-time dynamic programming algorithms such as Levenshtein's edit distance [1]. The main goal of these algorithms is to find out the minimum number of edits needed to make the read exactly match the reference segment [1]. Common edits include substitutions, insertions, and deletions. If the number of edits (called edit distance) is greater than a user-defined *edit distance threshold* (usually less than 5% of the read length [2-4]), then the mapping is considered to be invalid (i.e., the read does not match the segment at seed location) and thus is rejected. Calculating the edit distance for billions of sequences incurs significant computational burden [4-6]. Given that understanding complex diseases such as autism and cancer requires sequencing hundreds of thousands to millions of genomes [7, 8], the long execution time of today's read mappers can severely hinder such studies.

A wide variety of algorithms have been proposed to efficiently calculate the edit distance of sequences and filter out invalid mappings. Most existing algorithms can be divided into two main approaches: (1) accelerating the dynamic programming algorithms, (2) developing filtering heuristics that eliminate some of the invalid mappings (especially the ones that contain far more edits than allowed) before the verification step. Of the first approach, the classical dynamic programming algorithms such as Smith-Waterman [9], Levenshtein's edit distance [1], and Needleman-Wunsch [10] are the most accurate algorithms but they are computationally expensive as they require a quadratic running time. Subsequently, they were improved by computing only some necessary regions of the dynamic programming matrix rather than the entire matrix (e.g., Ukkonen [11]). They also can be accelerated by exploiting bit-parallelism in their implementations (e.g., Myers [12], SeqAn [13], SWPS3

Manuscript received June 15, 2017. This study is supported by NIH Grant (HG006004 to O. Mutlu and C. Alkan) and a Marie Curie Career Integration Grant (PCIG-2011-303772) to C. Alkan under the Seventh Framework Programme. M. Alser also acknowledges support from the Scientific and Technological Research Council of Turkey, under the TUBITAK 2215 program. M. Alser and C. Alkan are with the Computer Engineering Department, Bilkent University, 06800 Bilkent, Ankara, Turkey (e-mail: mohammedalser@bilkent.edu.tr, calkan@cs.bilkent.edu.tr). O. Mutlu is with the Computer Science Department, ETH Zürich, 8092 Zürich, Switzerland (e-mail: onur.mutlu@inf.ethz.ch).

[14], and hardware accelerated Smith-Waterman algorithm such as GPU-based [15] and FPGA-based [16]). The second approach to accelerate alignment verification is to incorporate a *filtering technique* within the read mapper and before the verification step. This filter is responsible for quickly excluding invalid mappings in an early stage (i.e., as a pre-alignment step) to reduce the number of locations that must be verified via dynamic programming. There are several existing filtering techniques such as Adjacency Filtering from FastHASH [6] and the Shifted Hamming Distance (SHD) [4].

We select SHD as the main focus of our analytical study, since it outperforms the Adjacency Filter in terms of speed and accuracy [4, 17]. It also maintains multiple independent bit-vectors (called shifted Hamming masks and explained in Section 2) that makes it suitable for parallel implementation. These filtering heuristics do *not* replace the verification step. Hence, they should be able to eliminate enough of the invalid mappings to be worthwhile (to compensate the computation overhead introduced by the filtering technique). One limitation with SHD is that it introduces inaccuracy in the filtering mechanism, allowing invalid mappings to pass the filter as false positives. A high number of false positives is undesirable, as these invalid mappings incur additional computational burden (they are unnecessarily examined *twice*, by *both* the pre-alignment and the alignment steps).

In this paper, our goal is to provide a detailed analysis of the false positive sources of the state-of-the-art alignment filter, SHD, aiming at eliminating them and boosting the performance of existing and future read mappers. To the best of our knowledge, this is the first paper to comprehensively assess the filtering inaccuracy of the SHD algorithm and provide recommendations for desirable improvements. The contributions of this paper are as follows:
- We provide a detailed investigation of four potential false positive sources of the state-of-the-art alignment filter, SHD.
- We show that processing the short matches (i.e., less than three matches) between two genomic sequences is not efficient, as they exhibit an unpredictable (random-like) and highly irregular behavior. Instead, future alignment filters should pay more attention to the *long, exact* matches shared by the sequences. Based on our observation, we build MAGNET, an intelligent filter that accurately detects all *long, exact* matches shared between two genome sequences.
- We quantify the false positives and true negatives of MAGNET and SHD using real data sets. We also experimentally demonstrate that incorporating long-match-awareness into the design of a pre-alignment filter can greatly improve the filtering accuracy.

## 2. OVERVIEW OF SHIFTED HAMMING DISTANCE

To provide a proper analysis of the false positive rate of SHD, in this section, we describe the SHD algorithm [4] and provide an example to illustrate how it works. SHD is a filter specifically developed to accelerate the alignment verification procedure in read mapping. SHD implements a filtering strategy that is inspired by the *pigeonhole principle*. That is, if $E$ items are put into $E+1$ boxes, then one or more boxes would be empty. This principle can be applied in the context of sequence alignment, as follows: if two reads differ by $E$ edits, then they should share at least a single identical subsequence (i.e., free of edits) among $E+1$ non-overlapping subsequences, where $E$ is the edit distance threshold. This is due to the fact that the $E$ edits would result in dividing the read into $E+1$ identical subsequences in accordance with their correspondences in the reference, as explained in Fig. 1. The more edits involved between two sequences, the less contiguous stretches of exact matches they share.

However, due to insertions and deletions, these identical subsequences might not be *perfectly* aligned and might be slightly shifted. Each insertion (or deletion) can shift multiple trailing bases to the right direction (or the left direction). SHD realigns the identical subsequences by incrementally shifting the read sequence against the reference sequence. SHD first calculates the base-pair-wise XOR between the two sequences. Then, it performs $E$ incremental shifts to the right direction to detect any read that has at most $E$ deletions, where $E$ is the edit distance threshold. Similarly, SHD also performs another $E$ incremental shifts to the left direction to detect any read that has at most $E$ insertions. After each shift, SHD calculates the base-pair-wise XOR between the read and the reference and stores the result in a shifted Hamming mask. In total, SHD generates $2E+1$ shifted Hamming masks regardless the source of the edit.

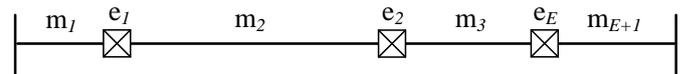

**Fig. 1: Random edit distribution in a read sequence. The edits ($e_1$, $e_2$, ..., $e_E$) act as dividers resulting in several identical subsequences ($m_1$, $m_2$, ..., $m_{E+1}$) between the read and the reference.**

Identical subsequences are then identified in each mask as a streak of continuous '0's. SHD ANDs all shifted Hamming masks together with the idea that all '0's in the individual Hamming masks propagate to the final bit-vector, thereby preserving the information of individual matching subsequences. SHD calculates the number of edits by counting the number of '1's in the final bit-vector. As SHD uses a bitwise AND operation, a zero at any position in the 2*E*+1 Hamming masks leads to a '0' in the resulting final bit-vector at the same position. Hence, even if some Hamming masks show a mismatch at that position, a zero in some other masks leads to a match ('0') at the same position. This tends to *underestimate* the actual number of edits and eventually causes some incorrect mappings to pass. To fix this issue, SHD proposes the so-called *speculative removal of short-matches* (SRS) before ANDing the masks, which flips short streaks of '0's in each mask into '1's such that they do *not* mask out '1's in other Hamming masks. The number of zeros to be amended (SRS threshold) is set by default to two. That is, bit streams such as 101, 1001 are replaced with 111 and 1111, respectively. Amending short streaks of '0's to '1's could cause correct mappings to be *mistakenly* filtered out, as it may produce multiple ones in the final bit-vector. To avoid this possibility, SHD always underestimates the number of edits from streaks of '1's. If there are four or three consecutive '1's in the final bit-vector, SHD counts them as a single edit. Thus, the total number of edits reported by SHD could be smaller than the real number of edits. In Fig. 2, we provide an example of a candidate alignment with all masks that are generated by the SHD algorithm. A segment of consecutive matches in one-step right-shifted mask indicates that there is a single deletion that occurred in the read sequence.

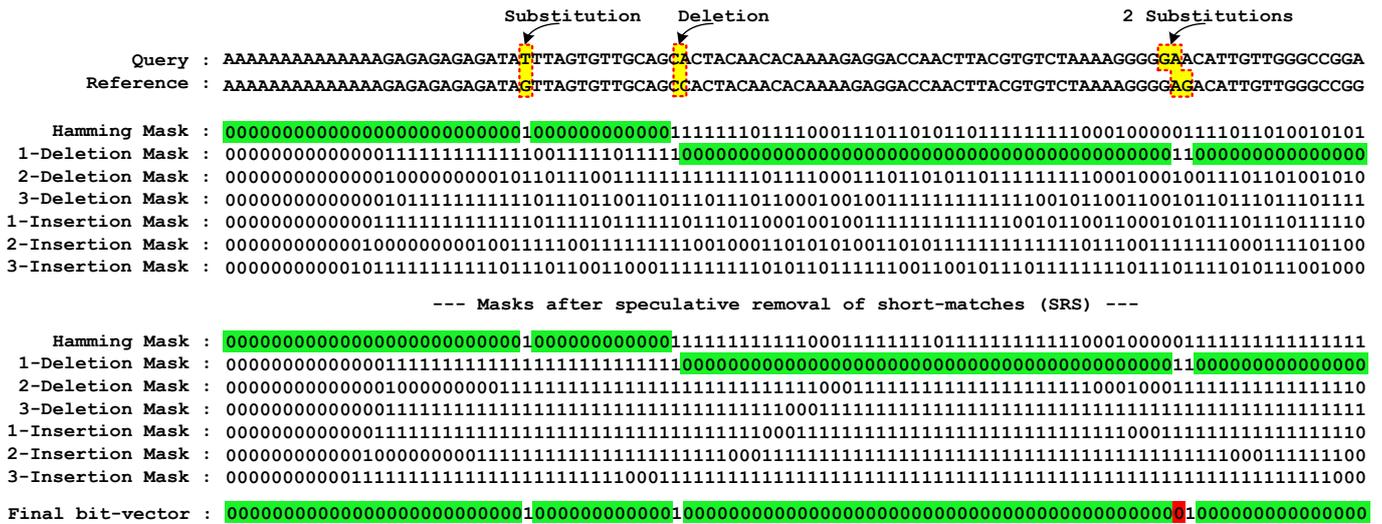

Fig. 2: An example of an alignment with all its generated masks, where the edit distance threshold (*E*) is set to 3. The green highlighted subsequences are part of the correct alignment. The red highlighted bit in the final bit-vector is a wrong alignment provided by SHD. The correct alignment (highlighted in yellow) shows that there are three substitutions and a single deletion, while SHD detects only two substitutions and a single deletion.

### 3. ON THE FALSE POSITIVES OF SHD

In this section, we investigate the potential sources of false positives that are introduced by the state-of-the-art filter, SHD [4]. We also provide examples that illustrate each of these sources of false positives. Adding an additional fast filtering heuristic before the verification step in a read mapper can be beneficial. But, such a filter can be easily worthless if it allows a high number of incorrect mappings to pass the filter. Even though the false positives that pass SHD are discarded later by the verification step (as the verification step has zero false positive rate), they can dramatically increase the running time of the read mapping by causing work to be done on a read by both the filtering step as well as the verification step. Below, we describe four major sources of false positives that are introduced by the filtering strategy of SHD.

#### A. Leading and Trailing Zeros

The first source of false positives in SHD is the streaks of zeros that are located at any of the two ends of each bit-vector. Hence we refer to them as leading and trailing zeros. These streaks of zeros can be in two forms: (1) the vacant bits that are caused by shifting the read against the reference segment and (2) the streaks of zeros that are not vacant bits. As we mentioned earlier, SHD generates *2E+1* masks using arithmetic left-shift and arithmetic right-shift operations. For both the left and right directions, the right-most and the left-most vacant bits, respectively, are filled with '0's. The number of vacant zeros depends on the number of shifted steps for each mask, which is at most equal to the edit distance threshold. The second form of the leading and trailing zeros is the zeros that are located at the two ends of the Hamming masks and are not vacant zeros. These streaks of zeros result from the pairwise comparison (i.e., bitwise XOR). They differ from the vacant bits in that their length is independent of the edit distance threshold. The main issue with both forms of leading and trailing zeros is that they always dominate, even if some Hamming masks show a mismatch at that position (due to the use of the AND operation). This gives the false impression that the read and the reference have a smaller edit distance, even when they differ significantly, as explained in Fig. 3. SRS does not address the inaccuracy caused by the leading and trailing zeros by amending such zeros to ones (as explained in Section 2), due to two reasons: (1) the number of these consecutive zeros is not fixed and thus they can be longer than the SRS threshold, (2) these consecutive zeros are *not* surrounded by ones and hence even if SRS threshold is greater than two bits, they are not eligible to be amended.

```
                    Vacant bits                                                                              Vacant bits
                    Leading zeros                                                                            Trailing zeros
       Query : AATCAAACAACCCCATCAACAAGTGGGCAAAGGATATGAACAGACACTTCTCAAAAGAAGACATTTATGCAGCCAACAGACACATGAAAAAATGCTCATC
   Reference : AAAAAACAACCCCATCAAAAAGTGGGCAAAGGATATGAACAGACACTTCTCAAAAGAAGACATTTATGCAGCCAACAGACACATGAAAAAATGCTCGTC
Hamming Mask : 00110000000000001000000000000000000000000000000000000000000000000000000000000000000000000000000000100
1-Deletion Mask : 001100011100011111111111111111111111111111111110001111111111111111111111111111111111111110000011111111
2-Deletion Mask : 001100011111111111111111111111111111111111111111111111111111111111111111111111111111111000111100001111111
3-Deletion Mask : 000100010001111111000111111111111111111111111111111100011111111111111111111111111111111100001111100
1-Insertion Mask : 001011111100011111111111111111111111111111111111111111111000111111111111111111111111111111100000111111110
2-Insertion Mask : 001111111111111111111111111111111111111111111111111111111111111111111111111111111111000111100001111111100
3-Insertion Mask : 001110001111111111111111111111111111111111111111111110000111111111111111111111111111111110001111100000

Final bit-vector : 0001000000000000000000000000000000000000000000000000000000000000000000000000000000000000000000000000

Needleman-Wunsch   AAAAAACAACCCCATCAAAAAGTGGGCAAAGGATATGAACAGACACTTCTCAAAAGAAGACATTTATGCAGCCAACAGACACATGAAAAAATGCTCGTC
      Alignment :    ||:||||||||||||||   |||||||||||||||||||||||||||||||||||||||||||||||||||||||||||||||||||||||||||||:||
                   AATCAAACAACCCCATCAACAAGTGGGCAAAGGATATGAACAGACACTTCTCAAAAGAAGACATTTATGCAGCCAACAGACACATGAAAAAATGCTCATC
```

**Fig. 3: Examples of an invalid mapping that passes the SHD filter due to the leading and trailing zeros (first source of false positives). We use an edit distance threshold of 3 and an SRS threshold of 2. While the regions that are highlighted in green are part of the correct alignment, the wrong alignment provided by SHD is highlighted in red. The yellow highlighted bits indicate a source of false positive.**

*B. Random Zeros*

The second source of false positives is the random zeros that appear in the individual shifted Hamming masks. Although they result from a pairwise comparison between a shifted read and a reference segment, we refer to them as random zeros because they are sometimes meaningless and are not part of the correct alignment. Different from the first source, these zeros are surrounded by ones and can be anywhere in the masks except the two ends of each bit-vector. However, the length and the position of these zeros are unpredictable. They can have any length that makes the SRS method ineffective at handling these random zeros. There is no clear theory behind the exact SRS threshold to be used to eliminate such zeros. SRS successfully removes some of the false positives, but it also introduces its own source of false positives.

Choosing a small SRS threshold helps, but does not provide any guarantee, to get rid of some of these random zeros. Choosing a larger SRS threshold can be risky, since, with such a large threshold, SHD might no longer be able to distinguish whether any streak of consecutive zeros is generated by random chance or it is part of the correct alignment. This results in SHD ignoring most of the exact matching subsequences and causes an all-'1' final bit-vector. In Fig. 4, we provide an example where random zeros dominate and lead to a zero in the final bit-vector at their corresponding locations. SRS can address the inaccuracy caused by the random 3-bit zeros, which are highlighted by the left arrow, using an SRS threshold of 3. However, SRS is still unable to solve the inaccuracy caused by the 15-bit zeros that are highlighted by the right arrow. This is due to the fact that the 15-bit zeros are part of the correct alignment and hence amending them to ones can introduce more false positives.

```
                 Random zeros
        Query : AAAAAAAAAAACCCATCAAAAAGTGGGCAAAGGATATGAACAGACACTTTTCAAAAGAAGACATTTATGCAGCCAAAAGACACATGAAAAAAATGCTCAT
    Reference : AAAAAAAAAACCCCATCAAAAAGTGGGCAAAGGATATGAACAGACACTTCTCAAAAGAAGACATTTATGCAGCCAACAGACACATGAAAAAATGCTCATC
 Hamming Mask : 00000000001000000000000000000000000000000001000000000000000000000000001000000000000001111111
1-Deletion Mask: 000000000111111100001111111111111111111111111110001111111111111111111111111110000011111111
2-Deletion Mask: 000000000111111100011111111111111111111111111111111111111111111111111111111000111100001111100
3-Deletion Mask: 000000000111111111111111111111111111111111111111111111000011111111111111111111111110001111111
1-Insertion Mask: 0000000000011100000111111111111111111111111110001111111111111111111111111110000000000000000
2-Insertion Mask: 000000000111110001111111111111111111111111110001111111111111111111111111111000111110001111100
3-Insertion Mask: 000000111111111111111111111111111111111111100001111111111111111111111111110000111111111000
Final bit-vector: 000000000000000000000000000000000000000000000000000000000000000000000000001000000000000000000000000
    Needleman-Wunsch AAAAAAAAAAACCCATCAAAAAGTGGGCAAAGGATATGAACAGACACTTTTCAAAAGAAGACATTTATGCAGCCAAAAGACACATGAAAAAAATGCTCAT
        Alignment: ||||||||||| ||||||||||||||||||||||||||||||||||||||  |||||||||||||||||||||||||||  ||||||||| |||||||||||||||
                  AAAAAAAAAACCCCATCAAAAAGTGGGCAAAGGATATGAACAGACACTTCTCAAAAGAAGACATTTATGCAGCCAACAGACACATG-AAAAATGCTCAT
```

**Fig 4: Examples of an incorrect mapping that passes the SHD filter due to random zeros (second source false positives). While the edit distance threshold is 3, a mapping of 4 edits (as examined at the end of the figure by Needleman-Wunsch algorithm) passes as a false positive.**

*C. Conservative Counting*

The third source of false positives is related to the way in which SHD counts the edits in the final bit-vector. As we discussed earlier, the amendment process can cause correct mappings to be mistakenly filtered out, as it may produce multiple '1's in the final bit-vector. To avoid this possibility, SHD counts the number of edits from any streak of '1's in the final bit-vector in a conservative manner. To ensure that it does not overcount the number of edits, SHD always assumes the streaks of '1's in the final bit-vector as a side effect of the SRS amendment, and counts only the minimum number of edits that potentially generate such a streak of '1's. The total number of edits reported by SHD can be much smaller than the actual number of edits. For instance, as illustrated in Fig. 5, three consecutive substitutions render a streak of three '1's in the final bit-vector. But since SHD always assumes the middle '1' is the result of an amended '0' by SRS, SHD will only consider the streak of three '1's as a single edit and let it pass, even if the edit distance threshold is less than three.

```
The 3-bit ones are a result of substitutions and not the amendment
        Query : AAAAAAACAAACAACCCCATCAAAAAGTGGGTGAAGGATATGAATTCACACTTCTCAAAAGAAGACATTTCTCAGCCAAAAACACATGAAAAAATGCTC
    Reference : AAAAAAACAAACAACCCCATCAAAAAGTGGGTGAAGGATATGAACAGACACTTCTCAAAAGAAGACATTTACTCAGCCAAAAAACACATGAAAAAATGCT
 Hamming Mask : 000000000000000000000000000000000000000000001110000000000000000000000001111111100000111111000001111
1-Deletion Mask: 00000001111111000111000011111111111111111111111111111111100011111111110000000000000000000000
2-Deletion Mask: 00000001111111111111100011111111111111111111111111110001111111111111111111111111100001111110000011
3-Deletion Mask: 000000111100011111111111111111111111111111111111111110001111111111110000100011110000001111
1-Insertion Mask: 00000011111110001110001111111111111111111111111111111110001111111111111000100011110000111100
2-Insertion Mask: 00000111111111111100011111111111111111111111111111111111110001111111110001111111111100011111100
3-Insertion Mask: 000111110001111111111111111111111111111111111111111110001111111111111111111111111111111111000
Final bit-vector: 000000000000000000000000000000000000000001110000000000000000000000001000000000000000000000000000
                                                            Misinterpreted as a single edit
    Needleman-Wunsch AAAAAAACAAACAACCCCATCAAAAAGTGGGTGAAGGATATGAATTCACACTTCTCAAAAGAAGACATTT-CTCAGCCAAAAAACACATGAAAAAATGCT
        Alignment: ||||||||||||||||||||||||||||||||||||||||||||   :|||||||||||||||||||||||| |||||||||||||||||||||||||||||||||
                  AAAAAAACAAACAACCCCATCAAAAAGTGGGTGAAGGATATGAACAGACACTTCTCAAAAGAAGACATTTACTCAGCCAAAAAACACATGAAAAAATGCT
```

**Fig 5: An example of an incorrect mapping that passes the SHD filter due to conservative counting of the short streak of '1's in the final bit-vector.**

*D. Lack of backtracking*

The last source of false positives in SHD is the inability of SHD to backtrack (after generating the final bit-vector) the location of each long identical subsequence (i.e., the mask that originates the identical subsequence), which is part of the correct alignment. The source of each subsequence provides a key insight into the actual number of edits between each two subsequences. That is, if a subsequence is located in a 2-step right shifted mask, it should indicate that there are two deletions before this subsequence. SHD does not relate this important fact to the number of edits in the final bit-vector. The lack of backtracking causes two types of false positives: (1) the first type of false positive in this category appears clearly when two of the identical subsequences, in the individual Hamming masks, are overlapped or nearly overlapped, (2) the second type happens when the identical subsequences come from different Hamming masks. The issue with the first type (i.e., overlapping subsequences) is the fact that they appear as a single identical subsequence in the final bit-vector, due to the use of AND operation. An example of this scenario is given in Fig. 6. This tends to hide some of the edits and eventually causes some invalid mappings to pass. The second type of false positives caused by the lack of backtracking happens, for example, when an identical subsequence comes from the first Hamming mask (i.e., with no shift) and the next identical subsequence comes from the 3-step left shifted mask. This scenario reveals that the number of edits between the two subsequences should not be less than three insertions. However, SHD inaccurately reports it as a single edit (due to ANDing all Hamming masks without backtracking the source of each streak of zeros), as illustrated in Fig. 7. Keeping track of the source mask of each identical subsequence prevents such false positives and helps to reveal the correct number of edits.

```
                     Overlapping subsequences can hide some edits
          Query : AAAAAAACAAACAACCCCAGAAAAAGTGGGTGAAGGACTATGAACAGACACTTCTCAAAAGAAGACTTTACTCAGCCAAAAAACACATGAAAAAATGCTA
      Reference : AAAAAAACAAACAACCCCATCAAAAAGTGGGTGAAGGATATGAACAGACACTTCTCAAAAGAAGACATTTACTCAGCCAAAAAACACATGAAAAAATGCT
   Hamming Mask : 000000000000000000011000011111111111110000000000000000000000000111111111110000011111110000011111
1-Deletion Mask : 00000001111111100011100000000000000001111111111111111000111111000000000000000000000000000000000
2-Deletion Mask : 000000011111111111100001111111111111111111111111111111111111111111000001111110000001111
3-Deletion Mask : 0000001110001111111000111111111111111111111111111111111100011111111111111100001000111100001111
1-Insertion Mask : 00000011111111000111000111111111111111111111111100011111111111111110000100011110000111110
2-Insertion Mask : 000011111111111111111111111111111111111111111111111111111111111110001111111110001111100
3-Insertion Mask : 000011111000111111111111110001111111111111111111111100011111111111111111111111111111111111000

Final bit-vector : 00000000000000000011000000000000000000000000000000000000000000010000000000000000000000000000000000

Needleman-Wunsch  AAAAAAACAAACAACCCCAG-AAAAGTGGGTGAAGGACTATGAACAGACACTTCTCAAAAGAAGAC-TTTACTCAGCCAAAAAACACATGAAAAAATGCT
     Alignment:   |||||||||||||||||||  |||||||||||||||    ||||||||||||||||||||||||| |||||||||||||||||||||||||||||||||
                  AAAAAAACAAACAACCCCATCAAAAAGTGGGTGAAGGA-TATGAACAGACACTTCTCAAAAGAAGACATTTACTCAGCCAAAAAACACATGAAAAAATGCT
```

**Fig 6: An example of an incorrect mapping that passes the SHD filter due to the lack of backtracking (overlapping identical subsequences).**

```
                                                      Backtracking this subsequence can tell
                                                      that it is a result of three insertions
          Query : AAAAAAAAAAATTAGCCAGGTGTGGTGGCACCCCCTGCCTATAATCCCAGCTACTCGGGAGGGAGGCAGGAGAATCGCTTGAACCTGGGAGGGGGAGGTT
      Reference : AAAAAAAAAAATTAGCCAGGTGTGGTGGCACATGCCTATAATCCCAGCTACTCGGGAGGCTGAGGCAGGAGAATCGCTTGAACCTGGGAGGGGGAGGTTG
   Hamming Mask : 000000000000000000000000000000001111111111111111111111111111111111111111111111111111111110000111111
1-Deletion Mask : 00000000001111111111111111111111000111111111111111111111111111111111111111111111111111111111110001111
2-Deletion Mask : 000000000011111111100011111111111111111111111111111111111111111111111111111111111111111111100011111
3-Deletion Mask : 000000001111111111000011111111111111111111111111111111111110001000111111111111111111111110001111110
1-Insertion Mask : 0000000001111111111111111111111111111111111111111111111110000000000000000000000000000000000
2-Insertion Mask : 00000000011111111100011111111111111111111111111111111111111111111111111111111111111111000011110000
3-Insertion Mask : 000000011111111111100011111110000000000000000000000000001111111111111111111111111111111000111111000

Final bit-vector : 00000000000000000000000000000000100000000000000000000000000011000000000000000000000000000000000000

Needleman-Wunsch  AAAAAAAAAAATTAGCCAGGTGTGGTGGCACCCCCTGCCTATAATCCCAGCTACTCGGGAGG--GAGGCAGGAGAATCGCTTGAACCTGGGAGGGGGAGGTT
     Alignment:   |||||||||||||||||||||||||||||||   ||||||||||||||||||||||||||||||  ||||||||||||||||||||||||||||||||||
                  AAAAAAAAAAATTAGCCAGGTGTGGTGGCAC---ATGCCTATAATCCCAGCTACTCGGGAGGCTGAGGCAGGAGAATCGCTTGAACCTGGGAGGGGGAGGTT
```

**Fig. 7: An example of an invalid mapping that passes the SHD filter due to the lack of backtracking.**

### E. Summary

We identify four sources that introduce the filtering inaccuracy of the SHD algorithm, namely, the leading and trailing zeros, random zeros, conservative counting, and lack of backtracking. Based on these four sources of false positives, we observe that there are still opportunities for further improvements on the accuracy of the state-of-the-art filter, SHD, which we discuss next.

### 4. MAGNET: OUR PROPOSED FILTERING STRATEGY

In this section, we first provide our own observations and recommendations based on our comprehensive accuracy analysis of the SHD filter. We then discuss our proposed filtering strategy, MAGNET, for genome analysis. Based on our analysis of the sources of false positives, we make two crucial observations.

The **first observation** is that handling the short streaks of '0's (i.e., using the SRS method that we discuss above) is indeed *inefficient*. These "noisy" streaks do *not* have determined properties, as their length and number are unpredictable (random-like). They introduce their own sources of false positives and do not contribute any useful information. Therefore, future filtering strategies should avoid processing such short streaks of '0's.

The **second observation** is that the correct (desired) alignment *always* contains all the longest non-overlapping identical subsequences. This turns our attention to focusing on the long matches (that are highlighted in green in all previous figures, i.e., Fig. 1 to Fig. 7) in each Hamming mask. We find that the long non-overlapping subsequences of consecutive zeros have three interesting properties. First, there is an upper bound on their quantity. With the existence of $E$ edits, there are at most $E+1$ non-overlapping identical subsequences (pigeonhole principle) shared between a pair of sequences. The total length of these non-overlapping subsequences is equal to $m-E$, where $m$ is the read length. Second, the length of the global longest subsequence is strictly not less than $[(m-E)/(E+1)]$. Third, the source mask of each long subsequence provides an insight into the number of edits between this subsequence and its preceding one.

These two observations motivate us to incorporate long-match-awareness into the design of our filtering strategy and ignore the short matches. MAGNET is a filtering heuristic that aims at finding all non-overlapping long streaks of consecutive zeros. By counting the number of these identical subsequences, we can infer the total number of edits between any pair of sequences (according to the first property that we discuss above). MAGNET algorithm consists of four main steps that can be explained as follows:

**Step 1:** MAGNET starts with searching for the first longest subsequence of consecutive zeros through all Hamming masks. It applies a sequential search algorithm along all $2E+1$ masks. Each mask nominates its local longest subsequence. Among all nominated subsequences, a single subsequence is selected as a global longest subsequence of zeros. Once found, our filter copies the target subsequence to the final bit-vector at the same corresponding location. It always attracts the longest subsequence of consecutive zeros and stores it in the final bit-vector and hence we call it MAGNET. All bits of the final bit-vector are initialized to '1'. The reason behind initializing it with '1's is that we want the final bit-vector to represent the number of mismatches.

**Step 2:** The next step is essential to preserving the original edit (or edits) that causes a single identical sequence to be divided into smaller subsequences. MAGNET penalizes the found subsequence by two edits (one for each side). This is achieved by excluding from the search space of all Hamming masks the indices of the found subsequence in addition to the index of the surrounding single bit from both left and right sides. So far we are able to track a single identical subsequence.

**Step 3:** In order to track the other non-overlapping subsequences, MAGNET applies a divide-and-conquer strategy where we decompose the problem of finding the longest identical subsequences into two subproblems. While, the first subproblem focuses on finding the next long subsequence that is located on the right-hand side of the previously found subsequence in the first step (i.e., Step 1), the second subproblem focuses on the other side of the found subsequence. Each subproblem is solved by recursively repeating all the three steps mentioned above. MAGNET applies an early termination method that aims at reducing the execution time of the alignment filtering by exploiting the first property of the long matches (i.e., the limited number of long matches).

```
       Query : AAAAAAAAAAAAAAGAGAGAGAGATATTTAGTGTTGCAGCACTACAACACAAAAGAGGACCAACTTACGTGTCTAAAAGGGGGAACATTGTTGGGCCGGA
   Reference : AAAAAAAAAAAAAAGAGAGAGAGATAGTTAGTGTTGCAGCCACTACAACACAAAAGAGGACCAACTTACGTGTCTAAAAGGGGGAGACATTGTTGGGCCGG
 Hamming Mask : 000000000000000000000000001000000000000001111110111100011011010110111111111000100000111101101001010 1
1-Deletion Mask : 000000000000001111111111111001111101111110000000000000000000000000000000000000110000000000000000
2-Deletion Mask : 000000000000001000000000101101110011111111111110111100011011010110111111111000100010011011010001010
3-Deletion Mask : 000000000000001011111111111011101100110011011101100010010011111111111000110110011 011011101111
1-Insertion Mask: 000000000000001111111111111011111011111101110110001001001111111111111000100010010010111011110111110
2-Insertion Mask: 000000000000001000000000100111110011111101001000110101010010101011111111111011100111110001111010100
3-Insertion Mask: 000000000000101111111111101101100110001111111101011011111001100101110111110110 1110101110010000

Final bit-vector : 00000000000000000000000001000000000001000000000000000000000000000000000000011000000000000000

Needleman-Wunsch   AAAAAAAAAAAAAAGAGAGAGAGATATTTAGTGTTGCAG-CACTACAACACAAAAGAGGACCAACTTACGTGTCTAAAAGGGGGAACATTGTTGGGCCGG
   Alignment:     ||||||||||||||||||||||||| ||||||||||||| ||||||||||||||||||||||||||||||||||||||||||::|||||||||||||||
                  AAAAAAAAAAAAAAGAGAGAGAGATAGTTAGTGTTGCAGCCACTACAACACAAAAGAGGACCAACTTACGTGTCTAAAAGGGGGAGACATTGTTGGGCCGG
```

**Fig 8: An example of the operation of our proposed filter, MAGNET. It shows the effect that incorporating long-matches-awareness has on the alignment accuracy. The alignment is compared to a sophisticated alignment algorithm (i.e. Needleman-Wunsch). Our algorithm finds all the longest non-overlapping subsequences of consecutive zeros in the descending order of their length (as numbered in yellow).**

Rather than searching for all long non-overlapping subsequences, our algorithm recursively solves the subproblems until the number of the subsequences found in the first step exceeds *E+1* or there are no more subproblems of size greater than or equal to a single bit. Each subproblem stores its solution (i.e., the longest identical subsequence) in the same final bit-vector that is shared by all subproblems.

**Step 4:** Once after the termination, MAGNET counts the occurrence of '1's in the final bit-vector. If their total number is equal or less than the edit distance threshold, *E*, then the mapping is considered to be valid. Likewise, if the total number of edits is sufficiently large (i.e. greater than a lower bound of edits), then the filter considers the mapping to be invalid and rejects it. In Fig. 8, we provide an example of how our filter works. Each '1' in the final bit-vector precisely reveals that there is an edit at its corresponding location of the Hamming mask.

With the help of our accuracy analysis of SHD (Section 3), we propose and incorporate long-match-awareness into the design of our filter. We get rid of the first three sources of false positives: (1) the leading and trailing zeros, (2) random zeros, and (3) conservative counting. In the next section, we investigate the impact of addressing these three sources on the false positive rate.

## 5. EVALUATION

In this section, we evaluate the false positive rate, true negative rate, and execution time of our proposed filter, MAGNET, against the best-performing previous filter, SHD [4]. As defined in previous work [17], the false positive rate is the fraction of incorrect mappings that are accepted by the filter out of all mappings, and the true negative rate is the fraction of incorrect mappings that are rejected by the filter out of all mappings. We always want to minimize the false positive rate and maximize the true negative rate. We implement both filters in MATLAB R2015b. We use a MATLAB implementation (nwalign [18]) of the Needleman-Wunsch algorithm [10] to benchmark the two filters as this algorithm has a false positive rate of 0%. We use a popular seed-and-extend mapper, mrFAST [19], to retrieve all potential mappings (read-reference pairs) from five sets, each containing about 4 million reads of length 100 base-pairs, from the 1000 Genomes Project Phase I [8].

**False Positive Rate.** In Fig. 9, we show the false positive rates of SHD and our proposed filter, across different edit distance thresholds. We configure mrFAST to generate from each read set the first half million read-reference pairs that have no more than 5 edits. On average, SHD produces a false positive rate of 20%, which is significantly higher (on average 20x) than that of our MAGNET filter.

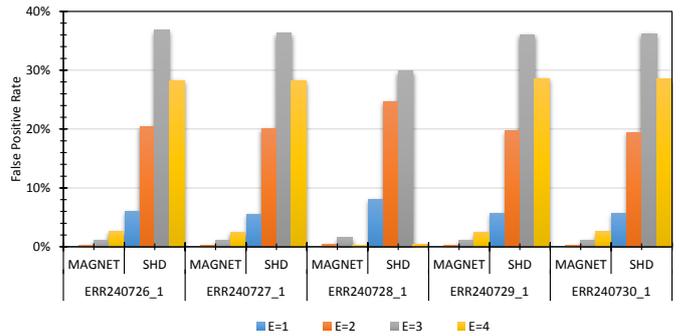

**Fig. 9:** The false positive rates of our MAGNET filter and SHD across different edit distance thresholds and read sets. The pairs are configured to have at most 5 edits.

In Fig. 10, we reconfigure mrFAST for an edit distance threshold of 7 (generated pairs have at most 7 edits). This enables us to measure the effectiveness of both filters when there are incorrect mappings that have a few more edits than the allowed threshold. We note that the number of false positives of SHD increases by at least 10%, while our filter still maintains a very low rate of false positives (<4%).

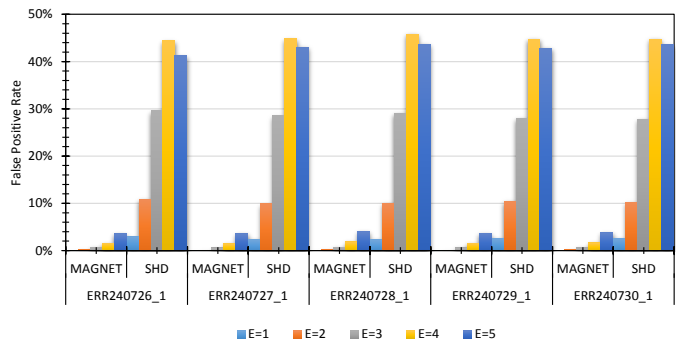

**Fig. 10:** The false positive rates of our MAGNET filter and SHD across different edit distance thresholds and read sets (configured to have at most 7 edits).

We now evaluate the false positive rate of MAGNET and SHD using the first 30 million pairs produced by mrFAST when the data set ERR240727_1 mapped to the human genome. We configure mrFAST to produce pairs that have at most 20 edits. Unlike the previous experiment, this configuration enables us to evaluate the false positive rate when the pairs have far more edits than the edit distance threshold. Fig. 11 demonstrates that SHD is more accurate in examining the edit-rich mappings than low-edit mappings. However, we find that MAGNET is very effective and superior to SHD in both situations (edit-rich and low-edit mappings). SHD falsely identifies potential mappings much more (15x - 100x, depending on the data and edit distance threshold used) than our filter.

We conclude that building an intelligent filter that is aware of all long matches is worthwhile and doing so significantly improves the accuracy of alignment filtering by at least an order of magnitude compared to the best previous filtering mechanism.

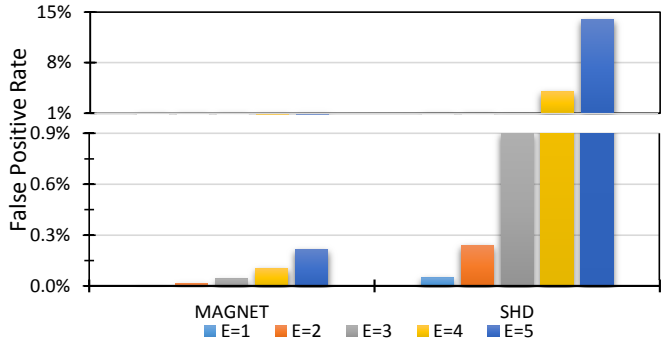

**Fig. 11: The false positive rates of MAGNET and SHD across different edit distance thresholds and using edit-rich mappings (having at most 20 edits).**

**True Negative Rate.** Next, we evaluate fraction of incorrect mappings that are rejected out of all rejected mappings, by both filters. We use in this experiment the first one million pairs that have at most 7 edits, produced by mrFAST when two data sets (ERR240726_1 and ERR240727_1) are mapped to the human genome. Fig. 12 shows that our filter rejects a significant fraction of incorrect mappings (e.g., up to 96%) and thus avoids expensive computations required by the verification step (dynamic programming). MAGNET rejects up to 20x more incorrect mappings than SHD. We conclude that our filtering strategy is more robust than SHD in handling invalid mappings. It boosts the overall performance by rejecting most of the incorrect mappings while at the same time providing a minimal false positive rate.

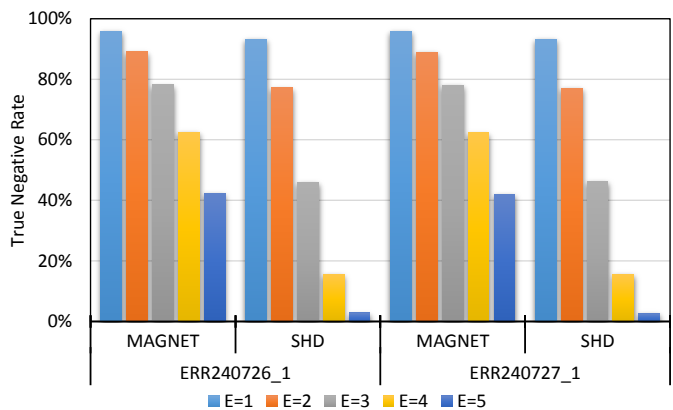

**Fig. 12: The true negative rates of MAGNET and SHD with different edit distance thresholds using one million mappings with at most 7 edits.**

**Execution time.** We now evaluate the execution time of our filter compared to the MATLAB implementation of the best existing filter, SHD, across different edit distance thresholds. We use the MATLAB Profiler [20] to track the execution time of both filters. We configure the Profiler to monitor the execution time based on the *performance clocking* option. Fig. 13 shows that as edit distance threshold increases, the execution time of both filters also increases. This is due to the fact that the number of Hamming masks is proportional to the edit distance threshold used and hence it requires more computations to be performed. We also find that MAGNET requires up to 1.4x more time than SHD to examine the first one million pairs that have at most 7 edits, produced by mrFAST when the data set ERR240726_1 is mapped to the human genome.

We conclude that our proposed filter, MAGNET, is extremely accurate, but this accuracy comes at the expense of a small increase in execution time. We believe this tradeoff is reasonable as examining the rejected mappings by a fast filter is much cheaper than having them verified by quadratic-time dynamic programming algorithms.

Note that the original SHD algorithm is implemented using Intel SSE instructions that provides significant performance improvements, especially for bit-parallel algorithms. We will explore in our future research how we can further improve the speed of our filtering strategy, to compete with the SIMD-implementation of SHD, using hardware accelerators (e.g., GPUs and FPGAs), multithreading, or SIMD instructions. In this work, we comprehensively evaluate the accuracy of the state-of-the-art alignment filter and mainly focus on addressing the sources of its filtering inaccuracy.

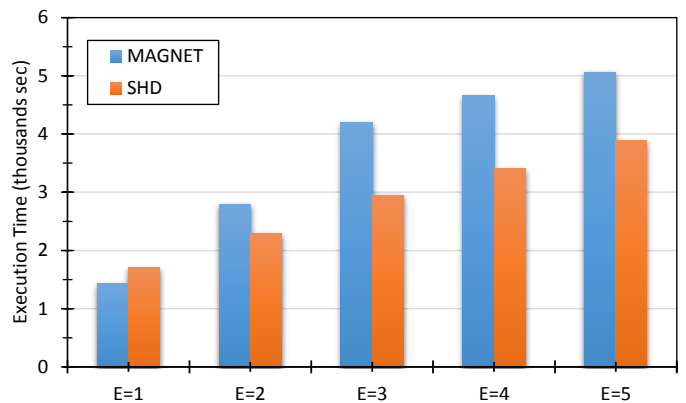

**Fig. 13: Execution time performance of MAGNET and SHD under different edit distance thresholds.**

## 6. CONCLUSION

In this paper, we comprehensively investigate four inaccuracy sources that make the state-of-the-art alignment filtering algorithm, Shifted Hamming Distance (SHD), highly ineffective in examining potential mappings in read mapping for genome analysis. We propose MAGNET, a new filtering strategy that eliminates these sources and significantly improves the accuracy of pre-alignment with a minimal false positive rate. In our experiments, we show that MAGNET correctly detects invalid mappings much better than SHD (i.e., we see reductions in the false positive rates as high as 15x - 100x, depending on the data and edit distance threshold used). We also show that MAGNET is able to reject up to 20x more incorrect mappings than SHD at the expense of a slight increase in the execution time. We believe that MAGNET is the most accurate pre-alignment filter in literature today, As such, we hope that our filtering strategy inspires researchers to adopt it and improve its implementation aiming at building an even faster yet extremely accurate pre-alignment filter.